\newcommand{\be}{\begin{equation}}
\newcommand{\ee}{\end{equation}}
\newcommand{\ber}{\begin{eqnarray}}
\newcommand{\eer}{\end{eqnarray}}

\documentclass[preprint,epsf,aps,floats]{revtex4}

\usepackage{psfrag}
\usepackage{epsfig}
\begin{document}
\tighten
\bigskip
\title{ 
Spontaneous Superfluid Current Generation in the kaon condensed CFL phase at Nonzero Strange Quark Mass}
\author{Andrei Kryjevski\footnote{akryjevski@physics.wustl.edu}}
\affiliation{Indiana~University,~Nuclear Theory Center,~Bloomington,~IN~47408 \footnote{Starting 09/07 the address is:
Washington University in St. Louis, Dept. of Physics, St. Louis, MO 63130}}
\date{\today}
\begin{abstract}
We find that for a large enough strange quark mass, $m_s^2/4\,\mu\,\Delta>2/3(1-0.023)$ ($\mu$ is the quark number chemical potential,
$\Delta$ is the superconducting gap), the kaon condensed CFL phase of asymptotically 
dense strongly interacting 3 flavor quark matter is unstable with respect to spontaneous generation of currents of Nambu Goldstone bosons 
due to spontaneous breaking of baryon number symmetry and hypercharge symmetry in the CFL$K^0$ ground state. The total baryon and 
hypercharge currents vanish in the ground state. We find that CFL$K^0$ and the new state are separated by a first order phase transition. 
The result is derived in the mean field approximation of High Density Effective Theory with electromagnetic interactions turned off.
\end{abstract}


\maketitle
\section{Introduction}
Properties of dense cold 
3 flavor quark matter characterized by a large quark number chemical potential $\mu\gg\Lambda_{QCD}$ and a low temperature 
${\rm T}\ll\Delta,$ where $\Delta$ is the superconducting gap, 
in the presence of non-zero quark masses have been a subject of an intensive investigation. See reviews 
\cite{Rajagopal:2000wf,Nardulli:2002ma,Schafer:2003vz,Shovkovy:2004me,Schaefer:2005ff,Alford:2006fw} and the references therein.~It has 
been argued that under the 
``stress'' induced by the strange quark mass, $m_s,$ the $SU(3)$ symmetric Color Flavor Locked (CFL) ground state of asymptotically dense 
quark matter \cite{Alford:1998mk} 
undergoes a second order phase transition to a less symmetric state where a condensate of collective bosonic excitations with quantum 
numbers of neutral kaons is formed (CFL$K^0$ phase) \cite{Bedaque:2001je,Kaplan:2001qk}. 
Later it was shown that the $K^0$ condensed ground state persists for larger values of $m_s,$ at least for
$y\equiv m_s^2/4\,\mu\,\Delta \leq 2/3$ 
\cite{Kryjevski:2004kt}. As the strange quark mass increases several quasiparticle excitations become lighter and at $y=2/3$ the spectrum 
of fermionic excitations develops a gapless 
electrically charged mode \cite{Kryjevski:2004kt,Kryjevski:2004jw}. It has been argued 
in the context of the symmetric CFL state that at the value of the strange quark mass where the gapless fermions appear ($y=1/2$), a phase 
characterized by flavor dependent quark-quark pairing pattern becomes favored (gapless CFL or gCFL phase) 
\cite{Alford:2003fq}. However, later it was
realized that because of gapless fermionic states in the spectrum, the proposed gCFL ground state had instabilities in the 
current-current correlation functions \cite{Casalbuoni:2004tb}. The resolution of these instabilities is currently under active 
investigation. Several instability free states have already been proposed 
\cite{Ciminale:2006sm,Gerhold:2006dt}.  

In this article, we present a calculation which suggests that for 
large enough strange quark mass, $m_s^2/4\,\mu\,\Delta>2/3(1-0.023),$ the CFL$K^0$ ground state of strongly interacting dense quark matter 
is unstable with respect to the generation of currents of Nambu Goldstone bosons (NGBs) due to spontaneous breaking of baryon number and 
hypercharge 
symmetry in CFL$K^0.$ A similar state in the context of a polarized Fermi gas near the unitary limit has been investigated in 
\cite{Son:2005qx}. Also, the idea of spontaneous NGB current generation in two flavor quark matter has been considered in 
\cite{Huang:2005pv,Hong:2005jv}.

This paper is organized as follows. In Section 2, we describe the assumptions and approximations made in this work, review the relevant 
low energy degrees of freedom and then perform a calculation 
of the free energy of the state with currents and find that it is favored for large enough $m_s.$ Section 3 contains conclusions and the 
outlook.

\section{Free Energy Calculation}
\subsection{Setup,  Approximations and Assumptions}
Let us start by listing the assumptions and approximations that we make in this work.

We will work in the isospin limit 
assuming $M={\rm diag}(m,m,m_s)$ for the quark mass 
matrix. Except for the Section 2.B. in this article, we will be neglecting small light quark mass, $m,$ as phenomenologically 
$m \ll m_s.$
We will assume that ${m_s^2/{\mu^2}} \ll 1$ and ${m_s^2/{4\,\Delta \mu}} \leq 1$
so that the light-strange pairing is still possible \cite{Schafer:1999pb}. 

In this regime the chiral expansion,
that is an expansion in ${\partial}/{\Delta}\sim{m_s^2/{\Delta \mu}},$ where one retains only few leading terms to 
attain the desired accuracy, is expected to break down and one should retain $m_s^2/{\Delta \mu}$ terms to all orders. The relevant 
effective theory in this regime should include both bosonic excitations of diquark condensate and quasiparticles as the degrees of freedom.

Being in the weak coupling regime (as $\mu\gg\Lambda_{QCD}$), we will work to leading order in $\alpha_s$ which corresponds to the mean 
field approximation. 

Also except for Section 2.B.,
we will neglect meson mass terms 
generated by quark masses. Quark mass terms connect particle and antiparticle states and the resulting meson mass terms are generally 
suppressed by powers of $\Delta/\mu\propto {\rm exp}(-{\rm const}/g_s)$ and/or $\alpha_s$ 
\cite{Son:1999cm,Schafer:2001za,Kryjevski:2004cw}. Also for simplicity we will neglect the small color-flavor symmetric CFL gap term as 
well as any other effects that may generate additional small mass gaps for quasiparticles (such as secondary gap generation due attractive 
meson exchanges). Investigation of such effects is left to future work.

For excitation energies below $\mu$ the relevant degrees of freedom are the nonet of quasiparticles 
and holes, ten pseudo NGBs due to the spontaneous symmetry breaking of global symmetries and soft gauge bosons. The corresponding 
effective theory is called High Density Effective Theory (HDET) \cite{Hong:1998tn,Hong:1999ru,Bedaque:2001je,Schafer:2003jn}.
Let us consider the leading terms of the HDET Lagrangian of the CFL phase of high density QCD in the mean field approximation
\ber
{\mathcal L} &=& 
- 2\times{{3 |\Delta|^2}\over{G}} + \nonumber \\ 
&+& {\rm Tr}\,\left[{L_v}^{\dag} 
i v \cdot D\,{L_v}\right] + 
{L_v}^{\dag}_{\,\,i\,a}\,
\left(i \vec{\sigma}_{\perp} \cdot \vec{D}\,\frac{1}{i\,\tilde{v} \cdot D+2\,\mu}\,i \vec{\sigma}_{\perp} \cdot \vec{D}\right)^a_b\,
{L_v}^{i\,b}
+ \nonumber \\  
&+& {{\Delta}\over{2}}\epsilon_{ijk}\epsilon_{abc}\left(\,e^{-2\,i\,\beta} X_{k\,c}\, {L^{ia}_{-v\,\alpha}}{L^{jb}_{v\,\rho}}\,
\epsilon^{\alpha\,\rho} + h.c.\right) +
\nonumber \\ &+& (L \leftrightarrow R,X \leftrightarrow Y,\mu_s^L \leftrightarrow \mu_s^R,\Delta \leftrightarrow -\Delta) -
\frac{1}{2}{\rm Tr}\,G^{\mu\nu} G_{\mu\nu}
- \frac{1}{4} F^{\mu\nu} F_{\mu\nu}.
\label{lagrangian1111}
\eer
Let us define the quantities in (\ref{lagrangian1111}) and list some additional approximations employed in this calculation.
\begin{itemize}
\item{
The first term is the mean field potential with $G \sim \alpha_s/\mu^2$ being the coupling of the effective 
attractive four fermion interaction generated by the hard gluon exchange \cite{Schafer:2003jn}.
}
\item{
The fermionic fields
$L_v^{i\,a}(x)$ and $R_v^{i\,a}(x)$ are the left- and right handed quasiparticle fields corresponding to states with Fermi velocity 
$v^{\mu} = (1,\hat v),\, \tilde v^{\mu} = (1,- \hat v),\,\hat v \cdot \hat v = 1;$ flavor indices
$i,j,k$ take values $u,d,s;$ color indices $a,b,c$ take values $r,g,b;$ 
$\alpha,\rho={\{}1,2{\}}$ are Weyl spinor indices (explicitly shown only in the gap term) and  
$\vec{\sigma_{\perp}}=\vec{\sigma}-\hat{v}\,(\vec{\sigma}\cdot\hat{v}).$ Chiral representation of $\gamma$ matrices is used. 
Trace over spin indices is not shown explicitly.
Under the original symmetry group $SU(3)_{c}\times SU(3)_{L}\times SU(3)_{R}\times U_B(1),$ 
$L_v$ transforms as $(3,3,1)_1$ and $R_v$ transforms as $(3,1,3)_1.$ 
}
\item{$Q={\mathrm diag}(2/3,-1/3,-1/3)$ is the quark electric charge matrix}
\item{
We have neglected the SU(3) singlet part of the leading HDET mass term, $M^2 /2\mu,$ which is responsible for the quark mass dependent 
baryon density shift, but does not affect the $m_s$ dependence of the condensation energy.
The relevant mass dependent terms are
\ber
\mu^L_s = \frac{M\,M^{\dag}}{2\mu} - \frac{1}{3}{\mathrm Tr} \frac{M\,M^{\dag}}{2\mu}, \nonumber \\
\mu^R_s = \frac{M^{\dag}\,M}{2\mu} - \frac{1}{3} {\mathrm Tr} \frac{M^{\dag}\,M}{2\mu}.
\label{mus}
\eer
}
\item{
The covariant derivatives for quark fields are defined as
\ber
&& D_{\nu} L = \partial_\nu L + i g_s L (A_\nu^{c})^{T} + i \left(e A^{em}_\nu Q + \delta_{\nu 0}\,{\mu^L_s}\right)L,\nonumber \\
   &&
 D_{\nu} R = \partial_\nu R + i g_s R (A_\nu^{c})^{T} + i \left(e A^{em}_\nu Q + \delta_{\nu 0}\,{\mu^R_s}\right)R,    
\label{cov_der}
 \eer
where $A_{\nu}^{c}$ and $A^{em}_{\nu}$ are the gluon and photon gauge fields, respectively; $A_{\nu}^c=A_{a\nu}^c t_a,$ where $t_a$ is an
$SU(3)_{c}$ generator in the fundamental representation.  It was argued in \cite{Bedaque:2001je} that ${M^2}/{
2 \mu}$ terms should appear in the covariant 
derivatives (\ref{cov_der}), and, thus, that in the low energy effective theory 
${m_s^2}/{2 \mu}$ plays the role of a chemical potential 
for strangeness.}
\item{
We consider the diquark condensate in the form
\ber
&&\langle{L_{ai}^*L_{bj}^*}\rangle\sim\epsilon_{kij}\epsilon_{cab}\,e^{-2\,i\,\beta}\,X_{k\,s}\,\Delta_{s\,c},\nonumber \\
&&\langle{R_{ai}^*R_{bj}^*}\rangle\sim-\epsilon_{kij}\epsilon_{cab}\,e^{-2\,i\,\beta}\,Y_{k\,s}\,\Delta_{s\,c}.
\label{gaptermXY}
\eer
The NGB fields are collected in $X_{kc}$ and
$Y_{kc}$ which are
$3 \times 3$ special unitary matrix valued composite 
fields describing oscillations of the left and right handed quark-quark condensates about the CFL ground state in the ``directions'' 
of spontaneously broken symmetries. 
Under the original symmetry group $SU(3)_{c}\times SU(3)_{L}\times SU(3)_{R}\times U(1)_B,$ 
$X\,\Delta$ transforms as $(3,3,1)_{-2}$ and $Y\,\Delta$ transforms as $(3,1,3)_{-2}$ 
\cite{Casalbuoni:1999wu,Son:1999cm}. In this calculation we will be assuming CFL pairing pattern and set 
$\Delta_{s\,c}=\delta_{s\,c}.$ 
Here $\beta$ is the (dimensionless) $U_B(1)$ NGB field.
The massive $U_A(1)$ pseudo-NGB field is neglected in this work.} 
\item{We use a mean field approximation in which both parts of the gap, 
$\Delta,$ and the chiral field, $\Sigma=\mathrm{X}\,\mathrm{Y}^{\dag},$ are to be determined from their equations of motion. 
For example, inclusion of quantum oscillations of the NGB fields about their ground state expectation values would give 
a contribution to the free energy density suppressed by ${\cal O}(\Delta/\mu)^2$ relative to the fermionic contribution.}
\item{
$G^{\mu\nu}$ and $F^{\mu\nu}$ are the field strength tensors for the gauge fields $A^{\mu}_{c}$ and $A_{em}^{\mu}.$ They vanish for the 
gauge field configurations relevant for this calculation.
}
\item{Also, we do not include oscillations of the gauge fields about their ground state expectation values.
These oscillations are important for the calculation of the gap parameter as the gap equation is dominated by the soft magnetic gluon 
exchanges \cite{Son:1998uk}, but here we do not attempt such a calculation. It was pointed by Weinberg that in the superconducting system
expression for the condensation free energy evaluated on the solution to the gap equation does not depend on the microscopic interaction 
between particles \cite{Weinberg:1993dw}.
Later Schaefer demonstrated that the expression for the free energy of quark matter in the mean field approximation when evaluated 
on the solution to the gap equation coincides with the expression for free energy of quark matter with color-flavor antisymmetric short 
range interactions in the mean field approximation (NJL model) \cite{Schafer:1999fe}.
With this in mind, our approximation 
is equivalent to working in the model with a color-flavor antisymmetric short range interactions (NJL model) in the mean field 
approximation with a given value of gap parameter \cite{Alford:1998mk,Schafer:1999fe} 
and the role of gauge fields is only to ensure the 
gauge charge neutrality of the ground state \cite{Alford:2002kj,Kryjevski:2003cu}.}

\end{itemize}

So, we consider dynamics of quasiparticles in some gauge charge neutral meson field background.
To calculate the free energy one needs to determine what this background is.

Changing basis of quasiparticle fields as
\ber
{L_v} &=& {1\over{\sqrt{2}}}\, e^{i \beta} \sum_{A=1}^{9}\,X\,\lambda^{A}\,{L^{A}_v},\nonumber \\
{R_v} &=& {1\over{\sqrt{2}}}\, e^{i \beta} \sum_{A=1}^{9}\,Y\,\lambda^{A}\,{R^{A}_v},
\label{q_chi}
\eer
where for $A=1,\ldots,8$ $\lambda^{ia}_{A}$ are the Gell Mann matrices (${\rm Tr}\,\lambda_A\lambda_B=2\delta_{AB}$) 
and $\lambda^{i\,a}_{9} = \sqrt {\frac{2}{3}} \,\delta^{i\,a}$ \cite{Bedaque:2001je}.
The Lagrangian (\ref{lagrangian1111}) may be rewritten as 
\ber
{\mathcal L} &=& - 
2\times{{3 |\Delta|^2}\over{G}}+ \nonumber \\
&&
\sum_{A,B=1}^{9} {\Psi_{L}^A}^{\dag}{\left( \begin{array}{ccc}
i\,v \cdot \partial\,\delta^{AB} + {\mathcal X}^{A\,B}_{\hat v}
& \Delta_A \delta^{AB} \\
\Delta_A \delta^{AB} & 
i\,\tilde{v}\cdot \partial\,\delta^{AB} - {\mathcal X}^{B\,A}_{-\hat v} \\
 \end{array} \right)} {\Psi_{L}^B} + \nonumber \\
&&
\sum_{A,B=1}^{9} {\Psi_{R}^A}^{\dag}{\left( \begin{array}{ccc}
i\,v \cdot \partial\,\delta^{AB} + {\mathcal Y}^{A\,B}_{\hat v}
& -\Delta_A \delta^{AB}  \\
-\Delta_A \delta^{AB}  & 
i\,\tilde{v} \cdot \partial\,\delta^{AB} - {\mathcal Y}^{B\,A}_{-\hat v}
 \\
 \end{array} \right)} {\Psi_{R}^B}, 
\label{action}
\eer
where $\Delta_A=\Delta,\,A=1..8$ and $\Delta_A=-2\,\Delta,\,A=9.$
The rest of the definitions are as follows:
\be
\Psi_{L}^A =
\left( \begin{array}{ccc}
L^A_v \\
(L^A_{-v})^{\dag} \end{array} \right)
\label{psiL}
\ee
and
\be
\Psi_R^A =
\left( \begin{array}{ccc}
R^{A}_v \\
(R^A_{-v})^{\dag} \end{array} \right)
\label{psiR}
\ee
are the components of the Nambu-Gor'kov field;

\ber
{\mathcal X}^{A\,B}_{v} &=& 
\sum_{C=1}^{9}{1\over{4}}{\rm Tr} \left(\,{\vec{\sigma}}_{\perp}\cdot i\,{\vec{\cal D}}_{A\,C}\,\left[X\,e^{i \beta}\right]\right)^{\dag} 
\frac{1}{i\,\tilde{v} \cdot \partial + 
2\,\mu}{\rm Tr} \,{\vec{\sigma}}_{\perp}\cdot i\,{\vec{\cal D}}_{C\,B}\,\left[X\,e^{i \beta}\right] +\nonumber \\ 
&+&
{1\over{2}}{\rm Tr}\,\lambda_{A} X^{\dag}\,e^{-i \beta} 
(i v \cdot \partial  - \mu_s^L - v \cdot A_{em}\,Q)\,e^{i \beta}\, X \lambda_{B} - 
{1\over{2}}\,{\rm Tr} \, \lambda_{A} \lambda_{B} \,v \cdot A_{c}^{T},
\label{JXYAB}
\eer
where 
\ber
i\,{\vec{\cal D}}_{A\,B}\left[X\right] =\lambda_{A}\,i 
\vec{\nabla}\,X\,\lambda_{B}-\lambda_{A}\,X\,\lambda_{B}\,{\vec A_c}^T.
\label{DAB}
\eer
The matrix ${\mathcal Y}^{A\,B}_{v}$ is given by the same expression as ${\mathcal X}^{A\,B}_{v}$ with $X$ and $Y$ interchanged. The gauge 
fields have been rescaled to eliminate explicit dependence on the gauge couplings. 

It has been shown that CFL 
is characterized by 
the following expressions for the (classical) gauge fields ({\it i.e.}, the following expressions solve the corresponding equations of 
motion) 
\cite{Casalbuoni:1999wu,Kryjevski:2003cu}
\ber
A_0^{c\,T} &=& -\frac{1}{2}(X^{\dag}(\mu_s^L + A^{em}_0 Q) X + Y^{\dag} (\mu_s^R + A^{em}_0 Q) Y),\nonumber \\
\vec{A_c}^T&=&-\frac{i}{2}\left(X^{\dag}\vec{\nabla}\,X + Y^{\dag}\vec{\nabla}\,Y\right).
\label{Asolution}
\eer
It was shown that for $M={\rm diag}(0,0,m_s)$ the CFL$K^0$ ground state is characterized by 
\cite{Bedaque:2001je,Kaplan:2001qk,Kryjevski:2004jw}
\ber
\Sigma &=&X\,Y^{\dag}= \Sigma_{K^0}=
\left( \begin{array}{ccc}
1 & 0 & 0 \\
0 & 0  & i \\
0 & i & 0 \end{array} \right),\nonumber \\
\vec{A_c}^T&=&0,\nonumber \\
A_0^{em}&=&0.
\label{sigmaK01}
\eer 

In the approximation used in this calculation the strange quark mass only appears in the role of an
effective chemical potential for strangeness. 
It has been argued that as $m_s$ increases the kaon mass terms generated by the quark masses suppressed by the powers of
$\Delta/\mu$ and/or $\alpha_s$ cannot counterbalance the ``chemical potential'', ${m_s^2/{2 \mu}},$ which is driving the $K^0$ Bose-Einstein
condensation \cite{Bedaque:2001je,Kaplan:2001qk}. For as long as the CFL pairing pattern is preferred,
there is nothing that can significantly alter the $K^0$ condensate in 
the ground state as $m_s$ increases. When the light quark masses are neglected there is literally nothing that can change the form 
of the kaon condensate in the ground state until a phase transition to a different ground state occurs \cite{Kryjevski:2004kt}.

Also it has been found that among fermionic excitations about the CFL$K^0$ ground state there is an electrically charged mode with a 
dispersion relation given by
\ber
\epsilon(p)/\Delta= -\frac{3\,y}{2} + \sqrt{\left(\frac{p}{\Delta}+\frac{y}{2}\right)^2 + 1}
\label{epsilon4}
\eer
with $p=\vec{p} \cdot \hat{v}-\mu.$ It becomes gapless at $y=m_s^2/4\,\mu\,\Delta=2/3$ \cite{Kryjevski:2004kt,Kryjevski:2004jw}.

\subsection{Low Energy Bosonic Excitations of CFL$K^0.$}

Before we proceed 
to the investigation of possible instabilities of CFL$K^0$ at large $m_s,$ 
let us recapitulate what the spectrum of collective bosonic excitations about the 
CFL$K^0$ ground state looks like \cite{Miransky:2001tw,Schafer:2001bq}. Here we work to the leading order in $M^2/\mu\,\Delta$ but will 
assume that the spectrum is qualitatively similar in the regime $m_s^2/4\,\mu\,\Delta \leq 2/3$ by continuity. For now let us reinstate 
the light quark masses so that 
$M={\rm diag}(m,m,m_s).$
We use the following parametrization for the pseudo-NGB excitations about CFL$K^0$ \cite{Kryjevski:2002ju}
\ber
\Sigma=\xi_{K^0}\,\hat {\Sigma}\,\xi_{K^0},
\label{SigmaK0param}
\eer
where $\hat {\Sigma}$ is the chiral octet field, $\xi_{K^0}^2=\Sigma_{K^0}.$ The superfluid mode terms are neglected here.
The leading terms of the chiral Lagrangian are
\ber
{\mathcal L} &=& 
{\mathrm Tr}\left[(\partial_t\,\hat{\pi})^2 - v^2\,(\vec{\nabla}\,\hat{\pi})^2 - 
i\,(\hat{\pi}\,(\partial_t\,\hat{\pi}) - (\partial_t\,\hat{\pi})\,\hat{\pi})\,(\mu_L+\mu_R) - a\,(M_L+M_R)\,\hat{\pi}^2 \right] + 
\nonumber \\ &+&
{\mathrm Tr}\left(b\,[Q_L,\hat{\pi}][Q_R,\hat{\pi}] -[\mu_L,\hat{\pi}][\mu_R,\hat{\pi}] \right),
\label{LK0}
\eer
where $M_L=\xi_{K^0}\,\bar{M}\,\xi_{K^0},\,$ $M_R=\xi_{K^0}^{\dag}\,\bar{M}\,\xi_{K^0}^{\dag},\,$ 
$\mu_L=\xi_{K^0}({M^2}/{2\,\mu})\xi_{K^0}^{\dag},\,$
$\mu_R=\xi_{K^0}^{\dag}({M^2}/{2\,\mu})\xi_{K^0},\,$ $Q_L=\xi_{K^0}^{\dag}\,Q\,\xi_{K^0},\,$ 
$Q_R=\xi_{K^0}\,Q\,\xi_{K^0}^{\dag},\,\bar{M}=|M|\,M^{-1}.$ 
The coefficients are as follows $a=3\,\Delta^2/\pi^2\,f_{\pi}^2,\,b\sim\,{\tilde \alpha}\,\Delta^2/4\,\pi,$ where ${\tilde \alpha}$ is 
the $U_{\tilde Q}(1)$ coupling constant 
\cite{Kaplan:2001qk,Son:1999cm}.
Let us first neglect the $\mathrm{b}$ term responsible for NGB mass splitting due to electromagnetic interactions between quarks. In the 
CFL$K^0$ state the 
$SU(2)\times U_Y(1)$ symmetry of the CFL phase is broken down to $U_{em}(1)\times P^{'}$ where $P^{'} = P\times Z,$ with 
$Z={\{}U_Y(2\,\pi\,\sqrt{3}),U_Y(4\,\pi\,\sqrt{3}){\}},$ the discrete 
symmetry of the $K^0$ condensed state. 

Three continuous symmetries are spontaneously broken. In the spectrum we find one massless mode with quadratic 
dispersion relation 
\ber
\epsilon^2_{K^{+},\, K^{-}}(k) = \left[\left(\frac{2\mu}{m_s^2}\right)^{2}-\frac{32\,a\,\mu^4\,m}{m_s^7}\right]\,v^4\,k^4 
\label{K1K2quadratic}
\eer
and one mode with linear dispersion relation  
\ber
\epsilon^2_{K^{0},\, \bar K^{0}}(k) = \left(1+\frac{64\,a^2\,m^2\,\mu^4}{m_s^6}\right)\,v^2 \, k^2,
\label{K01K02linear}
\eer
where $v^2=1/3,$ which is in agreement with the recent clarification to the Goldstone theorem on the counting of NGB modes 
\cite{Miransky:2001tw,Schafer:2001bq}. 
Finally, taking into account the $\mathrm{b}$ term responsible for electromagnetic mass splitting of the NGB masses gives the charged 
kaonic mode a mass $\mathrm{b}$ thus leaving two neutral massless NGB excitations of CFL$K^0$: the NGB due to spontaneous breaking of the 
hypercharge and the superfluid mode due spontaneous breaking of $U_B(1).$ Note that had we neglected light quark mass, we would have seen 
additional massless excitations in the pion-$\eta$ sector.

\subsection{Free Energy of the Current State}

Going back to the simplified $m=0$ case, we now investigate the stability of the CFL$K^0$ state against the spontaneous 
generation of currents made of these two massless collective bosonic excitations.

Let us work in the gauge
$X=\Sigma=X\,Y^{\dag},\,Y=1.$ A field configuration with a non-zero NGB field due to hypercharge spontaneously broken in CFL$K^0$ is
produced by applying the hypercharge transformation to $X = \Sigma_{K^0}$
\ber
X= \Sigma = U_Y\,\Sigma_{K^0}\,U_Y^{\dag}=
\left( \begin{array}{ccc}
1 & 0 & 0 \\
0 & 0  & i\,e^{-i\,\theta} \\
0 & i\,e^{i\,\theta} & 0 \end{array} \right),\,
Y=1,
\label{sigmaK01theta}
\eer
while a non-zero $\beta$ defined in (\ref{lagrangian1111}) describes a non-trivial $U_B(1)$ configuration.
   
Let us compare the free energies of the following two states: 1. CFL$K^0$ and 2. CFL$K^0$ with non-zero 
$\vec{j}_B\equiv\vec \nabla \beta$ and 
$\vec{j}_K\equiv\vec \nabla \theta$ present in the ground 
state. We will be making a simplifying assumption that $\vec{j}_k= const$ and $\vec{j}_B= const.$ The expression for the free energy 
difference is \cite{Weinberg:1993dw}
\ber
\delta\,{\Omega}&=& \Omega(\Sigma=\Sigma_{K^0},\Delta(j),\vec{j}_B,\vec{j}_K) - 
\Omega(\Sigma=\Sigma_{K^0},\Delta(0),\vec{j}_B=0,\vec{j}_K=0) = \nonumber \\
&=& 
\pi \int_{p} \sum_{A=1}^{9}\,
\left(\epsilon^{A}_L(p;m_s,\Delta(0),\vec{j}_B = 0,\vec{j}_K = 0) - \epsilon^{A}_L(p;m_s,\Delta(j),\vec{j}_B,\vec{j}_K \right)+\nonumber \\
&+&\pi \int_{p} \sum_{A=1}^{9}\,\left(\epsilon^{A}_R(p;m_s,\Delta(0),\vec{j}_B=0,\vec{j}_K = 0) - \epsilon^{A}_R(p;m_s,\Delta(j),\vec{j}_B,\vec{j}_K\right)+
\nonumber \\&+&
{3\,\mu^2\,|\vec{j}_B|^2}/{2\,\pi^2} +{3\,\mu^2\,|\vec{j}_K|^2}/{4\,\pi^2}.
\label{Omega_difference}
\eer
In equation (\ref{Omega_difference}):
\begin{itemize}
\item{the last two terms are the current kinetic terms which are discussed below (page 11)}
\item{${\{}p_0\pm\epsilon_{A}(p;m_s,\Delta(j),\vec{j}_B,\vec{j}_K){\}}$
are the eigenvalues of 
the inverse propagator (\ref{action}) for a given background, where $p_0$ is the 
time component of the 4-momentum vector} 
\item{The momentum integration is over the domain defined by 
$\epsilon_{A}(p;m_s,\vec{j}_B,\vec{j}_K) > 0.$
One has to exclude
contributions from the regions where $\epsilon_A < 0$ and to include contributions 
from the hole states where $-\epsilon_A > 0.$ {\it Note that a non-zero contribution to (\ref{Omega_difference}) may 
only come from the light quasiparticle mode which 
may actually become gapless due to the presence of the currents in the background.}} 
\item{the integration measure is 
\ber
\int_{p} = \frac{\mu^2}{(2\,\pi)^4}\int \,d\,p \,d \,\hat{v}.
\label{measure}
\eer}
\item{the value of the gap parameter may be different in the states with and without the currents, hence $\Delta=\Delta(j).$ But from now 
on we will be assuming $\Delta(0)=\Delta(j)\equiv \Delta.$ Validity of this assumption will be checked later.}
\end{itemize}

The last two terms in (\ref{Omega_difference}) are discussed in the next paragraph. Let us first determine the dispersion relation 
$\epsilon(p;m_s,\vec{j}_B,\vec{j}_K)$ for the light fermionic mode. 

After substitution  of (\ref{sigmaK01theta}) into (\ref{action})
the expressions for ${\mathcal X}_{v}$ and  ${\mathcal Y}_{v}$ of (\ref{JXYAB}) 
become
\ber
{\mathcal X}^{A\,C}_{v} &=& 
\left(- \hat{v} \cdot \vec{j}_B + i{\vec{\sigma}}_{\perp} \cdot \vec{j}_B\,\frac{1}{i\,\tilde{v} \cdot \partial + 2\,\mu}\,i 
{\vec{\sigma}}_{\perp} \cdot \vec{j}_B \right)
\delta^{A\,C}+\nonumber \\ &+& 
{1\over{2}}{\rm Tr} \, \lambda_{A}\,i{\vec{\sigma}}_{\perp} \cdot \vec{J}_K\,\frac{1}{i\,\tilde{v} \cdot \partial + 2\,\mu}\,i 
{\vec{\sigma}}_{\perp} \cdot \vec{J}_K\,\lambda_{C}+\nonumber \\ &+& 
{1\over{2}}{\rm Tr} \, \lambda_{A}\left(\hat{v} \cdot \vec{J}_K - X^{\dag}\,\mu_s^L\, X\,\right)\lambda_{C}
-{1\over{2}}\,{\rm Tr} \, \lambda_{A} \lambda_{C} \,(A_{c}^{0\,T} - \hat{v} \cdot \vec{A}_c^T)+...,
\label{JXABK}
\eer
\ber
{\mathcal Y}^{A\,C}_{v} &=& \left(- \hat{v} \cdot \vec{j}_B+ i{\vec{\sigma}}_{\perp}\cdot \vec{j_B}
\,\frac{1}{i\,\tilde{v}\cdot\partial + 2\,\mu}\,i 
{\vec{\sigma}}_{\perp} \cdot \vec{j}_B\right) 
\delta^{A\,C} +\nonumber \\ &-& {1\over{2}}{\mathrm Tr} \, \lambda_{A}\,\mu_s^R\, \lambda_{C} -
{1\over{2}}\,{\mathrm Tr} \, \lambda_{A} \lambda_{C} \,(A_{c}^{0\,T} - \hat{v} \cdot \vec{A}_c^T)+...,
\label{JYABK}
\eer
where 
\ber
\vec{J}_K
= X^{\dag}\,i\,\vec{\nabla}\,X =
\left( \begin{array}{ccc}
0 & 0 & 0 \\
0 & -1  & 0 \\
0 & 0 & 1 \end{array} \right)\,\vec{j}_K;
\label{JK}
\eer
the spatial gluon field induced by $\vec{j}_K,$
\ber
\vec{A_c}^T=-\frac{i}{2}\left(X^{\dag}\vec{\nabla}\,X + Y^{\dag}\vec{\nabla}\,Y\right)=\frac{1}{2}\,\left(\begin{array}{ccc}
0 & 0 & 0 \\
0 & -1  & 0 \\
0 & 0 & 1 \end{array} \right)\,\vec{j}_K,
\label{AcTi}
\eer
as well as the expression for $A_{c}^{0\,T}$ follow from (\ref{Asolution}) evaluated on (\ref{sigmaK01theta}).
We have neglected $j_B,\,j_K\sim{\cal O} (\Delta)$ terms compared to 
$2\,\mu$ in the denominator of the $1/(i\,\tilde{v} \cdot \partial + 2\,\mu)$ 
terms. Also, gauge field couplings in the numerator of the $1/(i\,\tilde{v} \cdot \partial + 2\,\mu)$ 
terms are not explicitly shown for simplicity. Their effect will be taken into account by gauge invariance.
Then, in the presence of $\vec{j}_B$ and $\vec{j}_K$ the light fermion dispersion relations are modified to
\ber
\epsilon_L(p,\hat{v}) &=& 
-\frac{3\,y}{2}\,\Delta +\hat{v} \cdot (\vec{j}_B + \frac{1}{4}\,\vec{j}_K) + \sqrt{\left(p + \frac{y}{2}\,\Delta -
\frac{3}{4}\,{\hat{v}.\vec{j}_K}\right)^2 + \Delta^2},\nonumber \\ 
\epsilon_R(p,\hat{v}) &=& 
-\frac{3\,y}{2}\,\Delta +\hat{v}\cdot (\vec{j}_B + \frac{1}{4}\,\vec{j}_K) + \sqrt{\left(p + \frac{y}{2}\,\Delta +
\frac{1}{4}\,{\hat{v}\cdot \vec{j}_K}\right)^2 + \Delta^2}.
\label{epsilon4betaK}
\eer
Here we have made an additional assumption that $A_0^{em}=0,$ which may not be true in the ground state we consider where charged 
quasiparticles are present. This issue will be discussed in the concluding section. 
We note that since (vacuum) parity is spontaneously broken in CFL$K^0,$ the excitation energies for the left and right handed species don't 
have to be equal. 
Near the minimum we approximate (\ref{epsilon4betaK}) as
\ber
\epsilon_L(p,\hat{v})&=&\epsilon_0\,\Delta + \hat{v} \cdot (\vec{j}_B + \frac{1}{4}\,\vec{j}_K) + \frac{(p-p_0^L)^2}{2\,\Delta},\nonumber \\
\epsilon_R(p,\hat{v})&=&\epsilon_0\,\Delta+ \hat{v} \cdot (\vec{j}_B + \frac{1}{4}\,\vec{j}_K) + \frac{(p-p_0^R)^2}{2\,\Delta},
\label{epsilon4betaapprLR}
\eer
where $\epsilon_0=1-3\,y/2,$ $p_0^L= - y\Delta/2 + 3\,{\hat{v}\cdot \vec{j_K}}/4,$ and 
$p_0^R=-y\Delta/2 - {\hat{v}\cdot \vec{j_K}}/4.$ 

{\it The following observation is of central importance for the calculation.} Eq. (\ref{epsilon4betaK}) indicates that in 
the presence of the currents the light fermionic mode may become gapless while 
$m_s^2/4\,\mu\,\Delta < 2/3.$ 
The dispersion curves dip below 0 when ${3\,y}/{2} -\hat{v} \cdot (\vec{j}_B + \frac{1}{4}\,\vec{j}_K) < 1;$ the domain defined by 
$\epsilon_{L,\,R}(p,\hat{v})>0$ is modified and the momentum integral in (\ref{Omega_difference}) has to be modified accordingly.

\begin{figure}[t]
\centering{
\begin{psfrags}
\psfrag{a}{$\mathrm{a})$}
\psfrag{b}{$\mathrm{b})$}
\psfrag{y}{$|\vec{j_B}|^2$}
\epsfig{figure=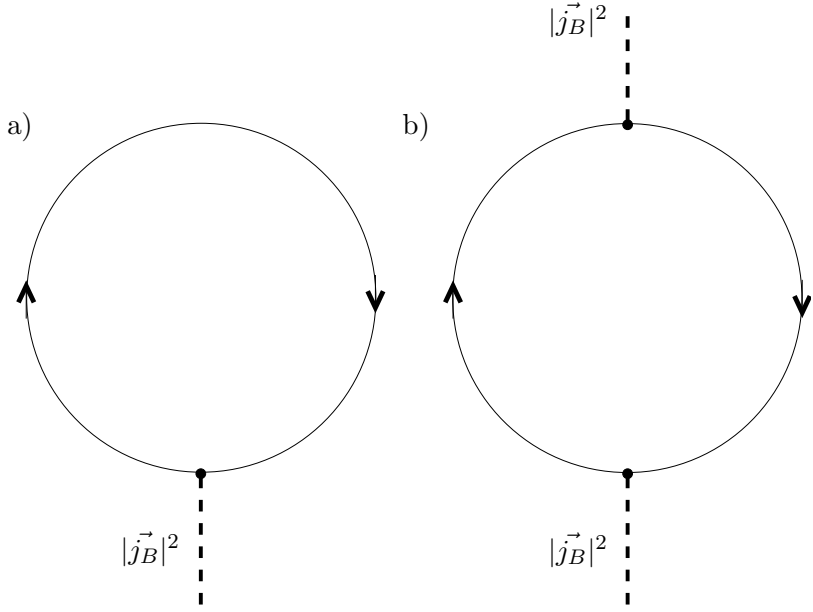, width=.65\textwidth}
\end{psfrags}
}
\caption{HDET graphs that produce normal contributions to the coefficients at the ${\cal O}(j_B^2)$ (Fig. a) and ${\cal O}(j_B^4)$ (Fig. b) 
terms in the effective Lagrangian. The ${\cal O}(j_B^4)$ terms are suppressed by two inverse powers of $\mu.$
}
\label{jB24}
\end{figure}

The last two terms in (\ref{Omega_difference}) are the 
kinetic term for $\beta(x)$ and the ``bare'' part of kinetic term for $\theta(x)$ of Eq. (\ref{sigmaK01theta})
(in terminology of Son and Stephanov \cite{Son:1999cm}).
They are generated by expanding the $1/(i\,\tilde{v} \cdot \partial + 2\,\mu)$ term in the effective action (\ref{action}) in 
$|\vec{j}_K|^2$ and in $|\vec{j}_B|^2.$ 
The ${\cal O}(j_{B,\,K}^2)$ terms are generated by a tadpole diagram with $|\vec{j}_B|^2,$ $\,|\vec{j}_K|^2$ or $\,|\vec{A}_c^T|$ insertion
into a quasiparticle loop (normal contribution to the $|\vec{j}_B|^2$ term is shown in Fig. \ref{jB24},a); the resulting loop integral 
needs to be regulated
and produces the ${\cal O}(\mu^2)$ coefficients shown in (\ref{Omega_difference})\cite{Casalbuoni:2000na,Son:1999cm}.  
The ${\cal O}(j_B^4),\,{\cal O}(j_K^4)$ terms are generated by a one loop HDET diagram with two $|\vec{j}_B|^2,\,|\vec{j}_K|^2$ insertions; 
the corresponding 
integral has two quasiparticle propagators and produces a coefficient ${\cal O}(\mu^0)$ which is suppressed by $1/\mu^2$ compared to the 
leading order term. Higher order 
terms are suppressed by higher powers of inverse $\mu.$ 
So, under the approximations used ({\it i.e.} to the leading order in $\mu,\, m_s^2/\mu^2$) only terms quadratic in $j_{B,\,K}$ are to be 
kept.


The expression for the free energy now reads 
\ber
\delta\,{\Omega}&=& \Omega(\Sigma=\Sigma_{K^0},\vec{j}_B,\,\vec{j}_K) - \Omega(\Sigma=\Sigma_{K^0},\vec{j}_B=0,\,\vec{j}_K=0) = 
\nonumber \\&=&
\frac{3}{2\,\pi^2}\,\mu^2\, |\vec{j}_B|^2 + \frac{21 - 8\,{\rm Log}\,2}{216\,\pi^2}\,\mu^2\,|\vec{j}_K|^2 
- \frac{\mu^2}{2(2\,\pi)^3}\int_{\hat{v}}\int_{p_m^L(\hat{v})}^{p_p^L(\hat{v})}d\,p\,\epsilon_L(p;\vec{j}_B,\,\vec{j}_K)- 
\nonumber \\
&-& \frac{\mu^2}{2(2\,\pi)^3}\int_{\hat{v}}\int_{p_m^R(\hat{v})}^{p_p^R(\hat{v})}d\,p\,\epsilon_R(p;\vec{j}_B,\,\vec{j}_K).
\label{Omega_difference_beta}
\eer
Here ${p_m^L(\hat{v})},\,{p_m^R(\hat{v})}$ and ${p_p^L(\hat{v})},{p_p^R(\hat{v})}$ are zeros of the gapless dispersion relation 
(\ref{epsilon4betaK}). The ``bare'' coefficient at the $j_K^2$ term has been augmented to the proper value for the $\theta(x)$ kinetic term 
previously found in the chiral perturbation theory 
for the CFL phase \cite{Son:1999cm}. The requisite contribution comes from 
\ber
\sum_{A=1}^{9}\,\pi\,\int_{\hat{v}}
\left[\int_{-\infty}^{\infty}d\,p\,\left(\epsilon_L^A(p;\vec{j}_B=0,\,\vec{j}_K=0)-\epsilon_L^A(p;\vec{j}_B,\,\vec{j}_K)\right)\right]+
(L\leftrightarrow R).
\label{restofjKkineticterm}
\eer

From now on we will be assuming that $\vec{j}_K$ and $\vec{j}_B$ are parallel, which obviously corresponds to the strongest effect. 
Performing the integration 
we arrive at the following expression for the free energy difference between CFL$K^0$ and the state with currents in the ground state 
(with $j_B$ and $j_K$ being in units of $\Delta$)
\cite{Son:2005qx}
\begin{eqnarray}
\frac{\delta\Omega}{\mu^2\,\Delta^2}&=&
-\Theta(|\vec{j}_B + \frac{1}{4}\,{\vec{j}_K}|-\epsilon_0)\,
\frac{2^{5/2}}{15\,\pi^2}
|\vec{j}_B + \frac{1}{4}\,{\vec{j}_K}|^{3/2}
\nonumber \\&\times&
\left(1-\frac{\epsilon_0}{|\vec{j}_B+ \frac{1}{4}\,{\vec{j}_K}|}\right)^{5/2}+
{f_{\beta}^2\,|\vec{j}_B|^2} +
\frac{f_{\pi}^2\,|\vec{j}_K|^2}{6}.
\label{Omega_difference_betaK}
\end{eqnarray}
Here $f_{\beta}^2={3}/{2\,\pi^2}$ and
$f_{\pi}^2={(21-8\,\rm{Log}\,2)}/{36\,\pi^2},$ $\epsilon_0$ is defined below (\ref{epsilon4betaapprLR}) 
and $\Theta(x)$ is a unit step function. 

\begin{figure}[t]
\centering{
\begin{psfrags}
\psfrag{E}{$\delta\,\Omega/\mu^2\,\Delta^2$}
\psfrag{jK}{$j_B/\Delta$}
\epsfig{figure=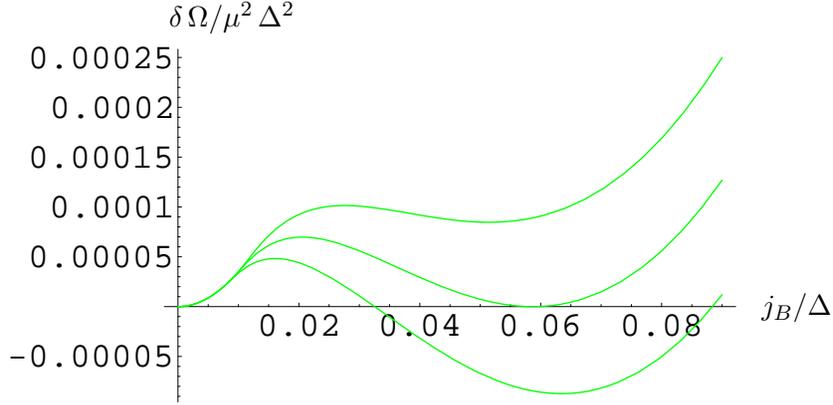, width=.65\textwidth}
\end{psfrags}
}
\caption{$\delta\,\Omega/\mu^2\,\Delta^2$ vs. $j_B/\Delta$ at $m_s$ corresponding to 
$\epsilon_0 \equiv 1-3\,y/2= {0.026,\,0.023,\,0.020}.$ 
}
\label{domega_BK1}
\end{figure}

Minimization of the free energy with respect to both currents
gives the relationship
\ber
|\vec{j}_K|=\frac{f_{\beta}^2}{24\,f_{\pi}^2}\,|\vec{j}_B| = \frac{81}{21-8\,\rm{Log}\,2}\,|\vec{j}_B|.
\label{jKjB}
\eer 
One gets the picture shown in (Fig. \ref{domega_BK1}). At $m_s^2/4\,\mu\,\Delta \simeq 2/3\,(1-0.023)$ the system undergoes 
a first order phase transition to a state with non-zero currents of Nambu Goldstone bosons due to spontaneous breaking of baryon number 
and of hypercharge symmetry in the CFL$K^0$ 
ground state with $|\vec{j}_B|=0.06\,\Delta$ and $|\vec{j}_K|=0.31\,\Delta.$

A more general calculation of the free energy difference where $\Delta(j) \neq \Delta(0)$ is allowed has been performed. Its results turned 
out to be very close to the results of the simplified calculation with $\Delta(j) = \Delta(0)$ thus justifying the 
earlier assumption. In this case the current terms are too small to appreciably modify the magnitude of the pairing gap.

The total baryon and hypercharge currents vanish in the ground state. A general way to see this is to recall that in the ground 
state the following equations of motion are satisfied
\ber
\delta \,{\rm L}_{eff}/\delta \vec{\nabla}\,\beta=0,\nonumber \\
\delta \,{\rm L}_{eff}/\delta \vec{\nabla}\,\theta=0,
\label{eomjBjK}
\eer 
where ${\rm L}_{eff}$ is the effective Lagrangian of the system. (Note that its exact form does not have to be specified in this argument.)
On the other hand, the corresponding Noether currents are given by 
the same expressions, and, therefore, have to vanish. More qualitatively, the NGB currents are 
canceled by the back flow of quasiparticles.

\section{Discussion and Outlook}

In this calculation we have used the simplest possible ansatz for the NGB currents: $\vec{j}_B=const,\,\vec{j}_K=const$ and 
found that 
in the vicinity of $m_s^2/4\,\mu\,\Delta=2/3$ where a light quasiparticle excitation is present in the spectrum, the CFL$K^0$ state is 
unstable with respect to spontaneous 
generation of baryon number and hypercharge NGB currents in the ground state. The two states are separated by a first order 
phase transition. Working to the leading order in $\alpha_s$ and neglecting ${\cal{O}}({m_s^2/{\mu^2}})$ terms we have found that the 
instability occurs for 
$m_s^2/4\,\mu\,\Delta \geq 2/3(1-0.023).$ 

Let us emphasize that strictly speaking the results obtained are valid in QCD with electro-magnetic interactions turned off. A more general
calculation of free energy for  $2/3 \leq m_s^2/4\,\mu\,\Delta\leq 1$
including effects of electric charge neutrality and the possibility of a more general flavor dependent quark-quark pairing pattern 
\ber
\langle{L_{ai}^*L_{bj}^*}\rangle\sim\epsilon_{kij}\epsilon_{cab}\,e^{-2\,i\,\beta}\,X_{k\,r}\,\Delta_{r\,c},
\label{noncflgap}
\eer 
with $\Delta={\rm diag}(\Delta_1,\Delta_2,\Delta_3)$\cite{Alford:2003fq}
will be presented elsewhere \cite{Kryjevski:2006ab}. Note that for $m_s^2/4\,\mu\,\Delta\leq 1$ a neutral 
quasiparticle mode is expected to become light \cite{Kryjevski:2004jw,Kryjevski:2004kt} which may further enhance the possibility of 
the current state formation.

Also, determination of the true ground state (even within a mean field approximation) requires solving the equations 
of motion for the NGB fields 
$\vec{j}_B(\vec{x})=\vec{\nabla} \cdot {{\beta}},\,\vec{j}_K(\vec{x})=\vec{\nabla} \cdot {\theta}.$ This is left to future work. It will 
be interesting to see if spatially inhomogeneous ground states suggested in the previous work 
on crystalline superconducting quark matter \cite{Alford:2000ze} will be found in this approach. 
Note that already in this simple calculation the gap term with non-zero $j_B$ and $j_K$ is essentially that of the 
Larkin-Ovchinnikov-Fulde-Ferrel (LOFF) phase with a single 
plane wave ansatz. 

Also, a non-zero kaon 
current generates a spatial gluon field in the ground state. Its role in a possible 
resolution of the chromomagnetic instability \cite{Casalbuoni:2004tb} is being investigated \cite{Kryjevski:2006ab}.

\begin{center}
\large{\textbf{Acknowledgments}}
\end{center}
I thank Dam T. Son for introducing me to the idea of spontaneous current generation in a superfluid system and helpful discussions. I 
also thank D.~Kaplan, M.~Stephanov and T.~Sch\"afer for discussions. I thank Andrew Crouse for proofreading the article. This
work is supported in part by the US Department of Energy grants DE-FG02-87ER40365 and DE-FG02-05ER41375 (OJI) and by the National Science 
Foundation grant PHY-0555232.

\bibliography{cfl}
%

\end{document}